\documentclass[aps,prd,twocolumn,twoside]{revtex4}
\pdfoutput=1
\usepackage{graphicx}
\usepackage{amsmath}
\usepackage{float}
\usepackage{amssymb}
\usepackage{caption}
\usepackage{placeins}
\usepackage{array}
\usepackage{subfig}
\usepackage{color}
\usepackage{pdfpages}
\newcommand{\comment}[1]{}

\def\vo{V$_\text{O}$ }
\def\vons{V$_\text{O}$}
\def\ah{$\alpha$-Fe$_2$O$_3$}
\def\ahs{$\alpha$-Fe$_2$O$_3$ }

\begin{document}
  \setcounter{page}{1}
  \date{\today}

\title{Effect of Defects on the Small Polaron Formation and Transport Properties \\ of Hematite from First-Principles Calculations}
\author{Tyler J. Smart$^1$, Yuan Ping$^2$}
\affiliation{$^1$Department of Physics, University of California-Santa Cruz, Santa Cruz, CA 95064 \\
$^2$Department of Chemistry, University of California-Santa Cruz, Santa Cruz, CA 95064}

\begin{abstract}
Hematite ($\alpha$-Fe$_2$O$_3$) is a promising candidate as photoanode materials for solar-to-fuel conversion due to its favorable band gap for visible light absorption, its stability in an aqueous environment and its relatively low cost in comparison to other prospective materials. However, the small polaron transport nature in $\alpha$-Fe$_2$O$_3$ results in low carrier mobility and conductivity, significantly lowering its efficiency from the theoretical limit. Experimentally, it has been found that the incorporation of oxygen vacancies and other dopants, such as Sn, into the material appreciably enhances its photo-to-current efficiency. Yet, no quantitative explanation has been provided to understand the role of oxygen vacancy or Sn-doping in hematite. We employed density functional theory to probe the small polaron formation in oxygen deficient hematite, N-doped as well as Sn-doped hematite. We computed the charged defect formation energies, the small polaron formation energy and hopping activation energies to understand the effect of defects on carrier concentration and mobility. This work provides us with a fundamental understanding regarding the role of defects on small polaron formation and transport properties in hematite, offering key insights into the design of new dopants to further improve the efficiency of transition metal oxides for solar-to-fuel conversion.
\end{abstract}

\maketitle

\section*{Introduction}

Several transition metal oxides, such as TiO$_2$, BiVO$_4$, WO$_3$ and hematite \ah , have exemplified some of the desirable characteristics of an efficient photoanode for solar-driven photoelectrochemical (PEC) water splitting \cite{photoanodes1,photoanodes2}. In particular, hematite (\ah) is a low cost, earth abundant, n-type semiconductor with a relatively smaller band gap in the visible range ($\sim\,$2 eV) compared with other oxides; thus it has been established as a promising candidate for photoanodes. However, application of hematite as a photoanode has been hindered from low carrier denstiy, low carrier mobility and high electron-hole recombination rate. Collectively, this results in the solar to fuel efficiency being substantially lower than the theoretical value, i.e. hematite typically yields a solar to hydrogen efficiency of 1-2\% despite having a theoretical efficiency for water splitting of 13\% \cite{efficiency}.

Extensive research has been conducted on the introduction of certain defects or dopants into hematite which could enhance PEC performance. In particular it has been shown that the introduction of defects such as \vons , Sn and Ti can improve the photocurrents of \ahs \cite{yatvo, yatsn, yatti}. Despite this, the underlying mechanism is not well understood. Specifically, whether the dopants improved the bulk properties or the interface charge separation, and how they affected light absorption, carrier concentration and carrier mobility of the bulk hematite, etc. Answering these questions is crucial for further optimization of chemical composition and morphology of hematite for higher PEC efficiency.

\begin{figure}[t!]
\begin{center}
\includegraphics[keepaspectratio=true,scale=0.25]{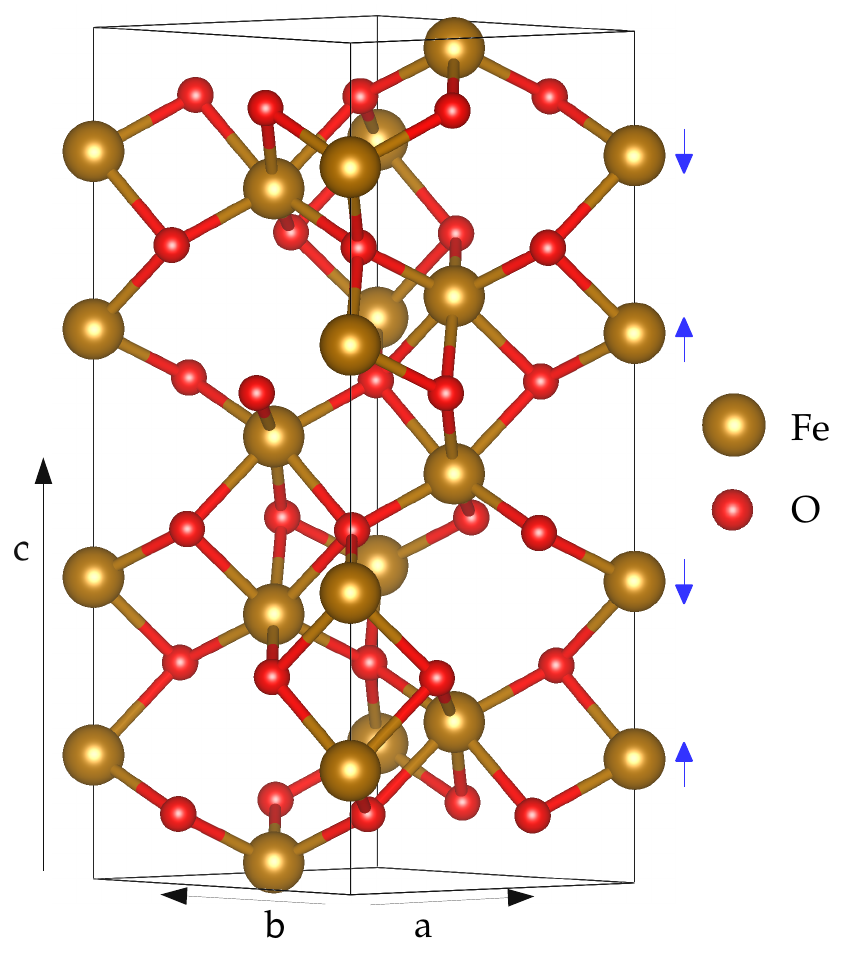}
\caption{The crystal structure of \ahs with lattice coordinates a, b and c. The blue arrows are to indicate the anti-ferromagnetic ordering of the iron ab layers.} \label{fig:pristine}
\end{center}
\end{figure}

Recently, the joint experimental and theory work by Ref.\citenum{ping} provided a detailed explanation for how nitrogen doping and oxygen vacancies in bismuth vanadate have led to improved photo-to-current efficiency by a simultaneous improvement of carrier density, carrier mobility and visible light absorption. In Ref.\citenum{ping}, it has been found that the enhanced mobility is tightly connected to the small polaron transport properties in BiVO$_4$, i.e. the N doping lowered the small polaron hopping barrier and improved the hopping mobility. Similarly, small polaron transport is the main carrier conduction mechanism in hematite, which causes its carrier mobility and conductivity to be extremely low. In the past, certain defects such as oxygen vacancies in bulk hematite have been discussed by computing their defect formation energies \cite{theoryvo,groupIV} but have not provided insights related to small polaron formation and conduction, which is the key to understand experimental observations (throughout this paper we will simply write polarons when referring to electron small polarons). 

Therefore we plan to illuminate the mechanism responsible for the enhancement of efficiency in doped hematite and provide a quantitative depiction of the roles of these dopants in terms of the small polaron formation and transport mobility in bulk hematite. We will discuss the electronic structure, defect charge transition levels and ionization energies for these individual defects as well as defect complex. Although several studies have been carried out on the formation energies of doped hematite \cite{theoryvo,groupIV}, their dependence on the choice of U parameters has not been examined in details. As we will show below, we have computed defect formation energies and ionization energies for different U parameters to investigate how the results vary as a function of U. Besides the defect ionization energies which are closely related to the carrier concentration, it is crucial to understand the effect of defects or dopants on the small polaron transport mobility in hematite which is another factor determining the carrier conductivity. To our best knowledge, discussion of small polaron transport in hematite from first principles is limited to pristine systems and there are few studies of small polaron transport in defective hematite \cite{trhematite}.In this work, we will examine how defects effect the small polaron hopping barriers for carrier conduction in doped hematite.
By this means we can answer important experimental questions related to the effects of defects on hematite and also provide extending insights for other doped transition metal oxides, supplying guidance for future experimental design of improved small polaron transport properties in metal oxides.

\section*{Methods}
We obtained the electronic structure of pristine and doped hematite with Density Functional Theory (DFT) calculations including Hubbard U corrections in order to take into account of strong on-site d electron interactions. Total energy, geometry relaxation and electronic structure calculations were performed with open source, plane wave codes Quantum ESPRESSO \cite{qe}. We used ultrasoft pseudopotentials \cite{GBRV} including Fe $3s$ and $3d$ semicore electrons,  which in our case yielded identical results to norm-conserving alternatives \cite{ONCV}. For the plane wave basis we individually used cutoff energies of 40 Ry for wave functions and 240 Ry for charge density. The Brillouin zone was sampled with a $4\times4\times2$ Monkhorst-Pack $k$-point mesh for geometry optimization of the 30 atom unit cell and the $k$-point mesh was doubled in density of states calculations. Methfessel-Paxton first-order spreading was used to expedite Brillouin zone integration with a smearing width of 0.001 Ry \cite{smear}. Total energy was calculated self consistently until a convergence of $10^{-8}$ Ry was achieved and geometry was optimized until the net force per atom reached less than 10$^{-3}$ Ry/au. For the hopping activation barrier calculations, a linear extrapolation scheme of atomic structure between initial and final hopping centers was employed for pristine and defective systems.\\
	
\ahs is arranged in the hexagonal corundum structure with a space group of R$\overline{3}$c \cite{twenties} (Figure \ref{fig:pristine}). The unit cell consists of 30 atoms where iron, Fe$^{3+}$, is six coordinated and oxygen, O$^{2-}$, is four coordinated. The system is anti-ferromagnetic where the Fe$^{3+}$ ions have a high spin configuration of partially occupied $3d$ orbitals with aligned spin between ions within $ab$-planes and anti-aligned along the $c$ direction. We first optimized the initial geometry of pristine hematite with a variable cell relaxation which obtained cell parameters of $a=5.13\,${\AA} and $c=13.99\,${\AA} with bond lengths and angles of Fe-O$\ = 1.99,\, 2.14\,${\AA}, O-Fe-O$\ = 90.7,\,86.0,\,78.6\, ^\circ$ , which agree well with other work in both experiment \cite{cell1} and theory \cite{transport}. The pristine structure and fundamental values that we obtained are collected in Table \ref{table:pristine}.\\

\begin{table}[H]
\begin{center}
\begin{tabular}{p{2.75cm}p{2.75cm}p{2.75cm}}
 \hline
 \multicolumn{3}{c}{ Pristine Hematite Fundamental Values \vspace{0.5mm}} \\ 
 \hline
 Value & Present Work & Experiment \cite{cell1} \\
 \hline
 $a$ ({\AA}) & 5.13  & 5.04 \\
 $c$ ({\AA}) & 13.99  & 13.75 \\
 Fe-O ({\AA}) & 1.99, 2.14  & 1.94, 2.11 \\
 O-Fe-O ($^{\circ}$) &  90.7, 86.0, 78.6 & 90.5, 85.0, 78.2 \\ 
 $\mu_{Fe}$ ($\mu_B$) & 4.0 & 4.6 \cite{magmom} \\
 $E_{gap}$ (eV) &  2.21  & 2.2  \cite{gap}\\
 \hline
\end{tabular}
\caption{Computed and experimental lattice constants $a$ and $c$ ($b=a$), nearest neighbor bond lengths Fe-O, bond angles O-Fe-O and the band gap energy $E_{gap}$.} \label{table:pristine}
\end{center}
\end{table}
	
Hematite is a charge transfer insulator \cite{ctinsulatorexp,ctinsulatorthe} and in accordance with the strong correlation of electrons in $3d$ orbitals we apply the Hubbard correction \cite{hubbard} to the Fe $3d$ orbitals in the form of Eq.\ref{eq:U}. Here, $E_\text{DFT}$ is the energy obtained from standard Density Functional Theory methods which is corrected by the following term which includes the occupation matrix $\lambda_i^{I\sigma}$ ($I$ ranges over all Fe ions, $i$ ranges over $3d$ orbitals and $\sigma$ is for spin up or down).
 \begin{equation}
E_{\text{DFT}+\text{U}}=E_{\text{DFT}}+\frac{U_{\text{eff}}}{2}\sum_{I,\sigma}\sum_{i}\lambda_{i}^{I\sigma}(1-\lambda_{i}^{I\sigma}) \label{eq:U}
\end{equation}


\begin{figure*}
\begin{center}
\includegraphics[keepaspectratio=true,scale=1]{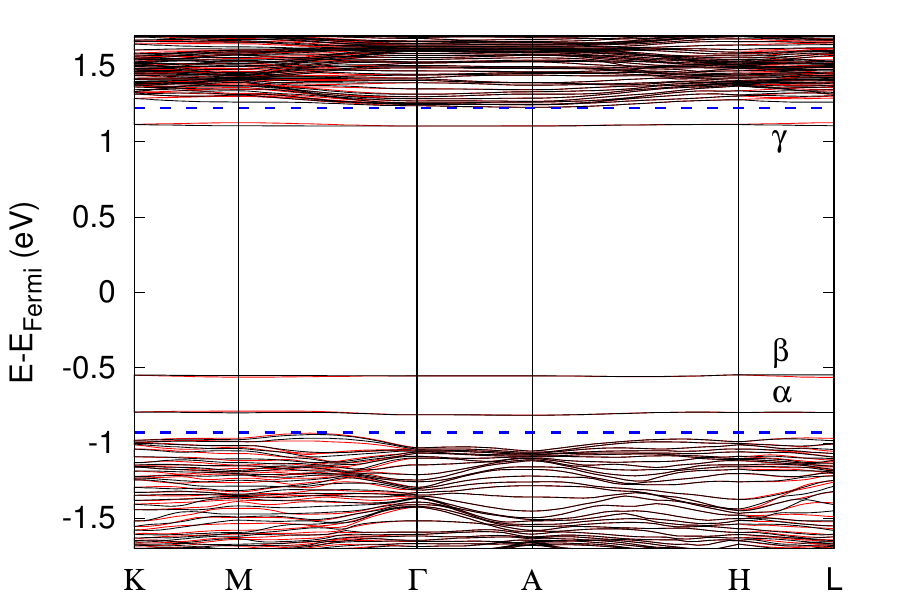} 
\vspace{2mm}
\includegraphics[keepaspectratio=true,scale=0.15]{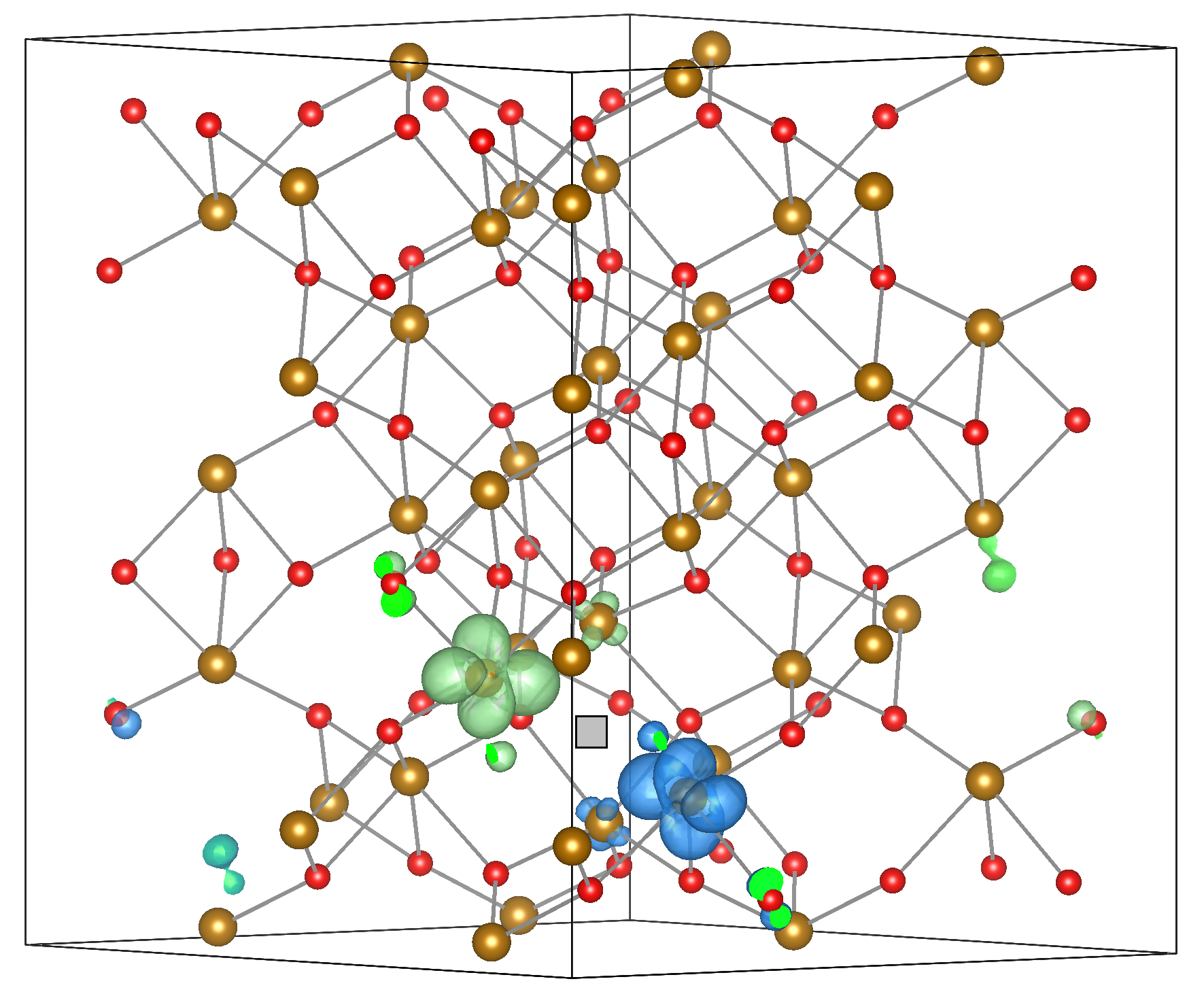}
\caption{(Left) The band structure of Vo:Fe$_2$O$_3$ with unperturbed band edges and the creation of three isolated defect states $\alpha$, $\beta$ and $\gamma$ (red curves: spin up; black curves: spin down; blue dashed lines mark VBM and CBM). Defect states are states which are away from the band edges and correspond to localized states as in this case. (Right) Isosurface plot of the two highest occupied orbitals present in Vo:Fe$_2$O$_3$ with an isosurface of $\sim\,$2\% the maximum value. These two degenerate small polarons form at nearest neighbor Fe of \vo (grey box) and are referred to as $\beta$ polarons throughout this paper (green isosurface spin up and blue isosurface spin down).} \label{fig:vo}
\end{center}
\end{figure*}


The $U_{\text{eff}}$ parameter, which sets the strength of this correction, was determined empirically to be 4.3 eV, in order to best fit the band gap of hematite to be 2.2 eV as measured by soft x-ray spectroscopy \cite{gap}. It is notable that Hubbard parameters used in previous calculations of this material vary, although most used a $U_{\text{eff}}$ between 4 eV to 4.3 eV \cite{theoryvo,transport,nativedefects,groupIV,theoryu,theoryu2,trhematite}. In particular, one study also used a non-empirical method, on the basis of unrestricted Hartree Fock theory, to determine a converged value of $U_{\text{eff}} = 4.3$ eV for \ah , which they also showed to provide results in good agreement with experiment \cite{theoryu}. \\

	
The introduction of any extra electrons into \ahs will form strongly localized small polarons due to large electron phonon couplings (the electrons are self-trapped by local lattice distortions) and the small polaron radius is typically no farther than the next-nearest neighbors \cite{polaron}. The formation of small polarons have been found in several metal oxides, such as Fe$_2$O$_3$, BiVO$_4$ \cite{ping}, and ABO$_{3}$ perovskites \cite{abo3}, where the small polarons must be thermally activated and then hop between metal sites in order to conduct currents. In this work we plan to study the relative stability and hopping barriers between different polaron configurations, where controlling the position of small polaron formation with/without defects is the key. We controlled the placement of the small polaron through local geometry adjustments that break the symmetry of the ground state system, such as moving the oxygen atoms surrounding an Fe ion away from it to ensure localization of polarons at that site and then relaxing the geometry to energy minimum. This bias initiates the localization of polarons which will expand the local metal-oxygen bonds slightly ($\sim$0.1 \AA).  The small polarons form both in pristine systems (without explicit defects) and defective systems. For the latter, the small polaron and defect center often interact and possibly form a complex as we will discuss later. 

For calculations with defects and excess charge, a $2\times 2\times 1$ supercell with a $2\times2\times2$ $k$-point mesh was adopted and internal geometries were relaxed keeping cell parameters fixed to simulate an isolated defect within the periodic system.  We found the $2\times 2\times 1$ supercell is large enough to avoid periodic interactions between neutral defects. The charged defect formation energies have a slow convergence with supercell size (scale as 1/L) due to image charge interactions. We corrected the charge defect formation energies using the charge correction scheme similar to Ref. \citenum{ccc} and our recent development in Ref. \citenum{ping2} where the proper treatment for electrostatic potential alignment at presence of strong geometry relaxation by defects is included, which has been a general issue for charged defect calculations \cite{Kumagai2014}.

\subsection*{Electronic Structure of Defective Hematite}

\subsubsection*{Oxygen Vacancy}

Oxygen vacancy is an important intrinsic defect in \ah , whose concentration can be controlled through experimental conditions. Recently, it has been found that the introduction of oxygen vacancies in 
\ahs can improve the overall photocurrents of hematite based photoanodes in solar water splitting cells \cite{yatvo}. Yet, the role of oxygen vacancies is unclear, and two possible explanations are present: (1) oxygen vacancies improve the bulk carrier conductivity properties or (2) oxygen vacancies improve the electron hole separation at electrode/electrolyte interfaces. In this work we focus on (1), i.e. the effect of oxygen vacancies on the bulk properties of \ah.


\begin{figure*}
\begin{center}
\includegraphics[keepaspectratio=true,scale=1]{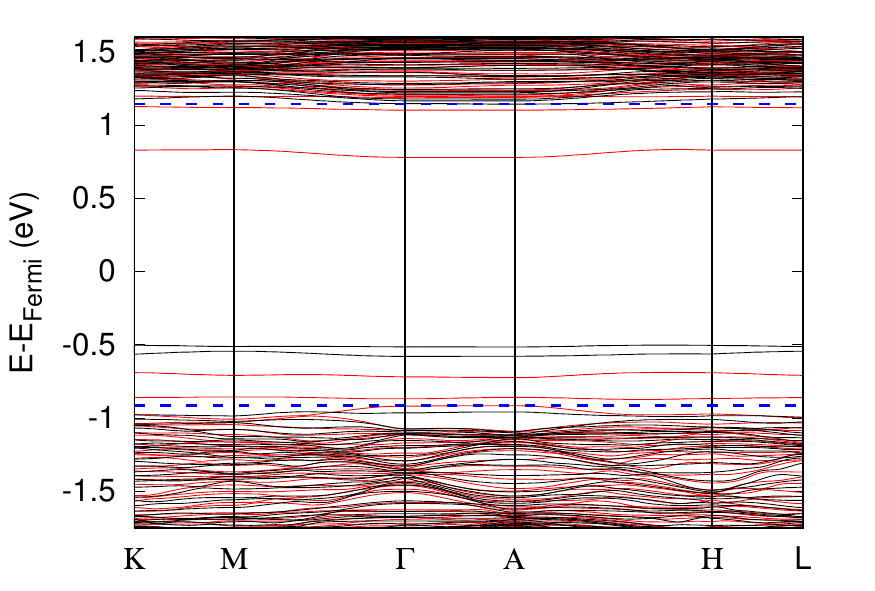}
\includegraphics[keepaspectratio=true,scale=0.15]{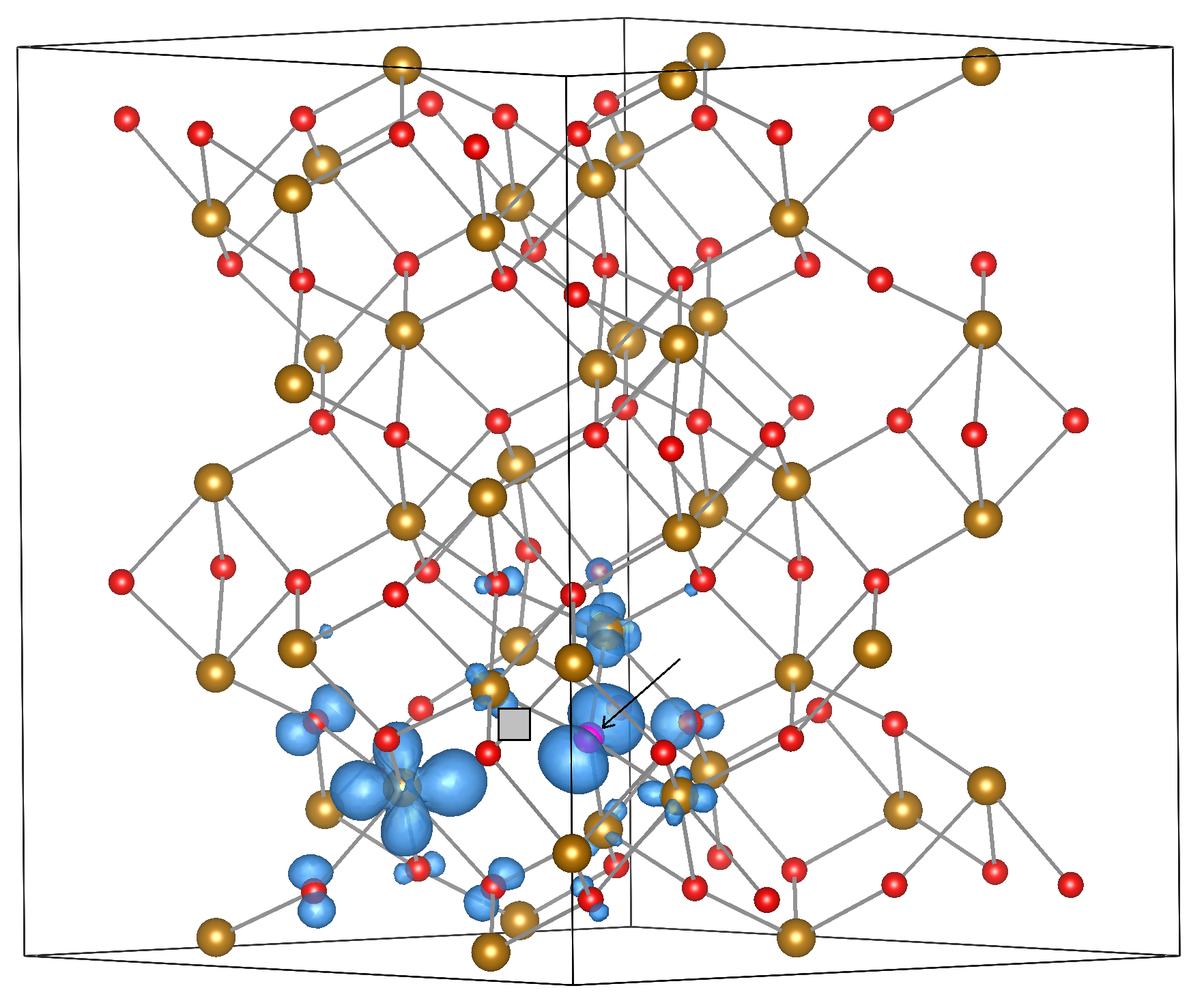}
\caption{(Top) The band structure of N+Vo:Fe$_2$O$_3$ with a perturbed valence band edge (right below the lower dashed blue line), creating a reduced gap and an indirect to direct gap transition (red curves: spin up and black curves: spin down; blue dashed lines mark VBM and CBM). There are several defect states introduced in the gap (between two blue dashed lines), including occupied defect states which are primarily hybridized Fe $3d$ and N $2p$ states. (Bottom) Isosurface plot (blue) of the highest occupied orbital present in N+Vo:Fe$_2$O$_3$, a combination of a $\beta$ polaron from the presence of \vo (grey box) and N $2p$ orbital (pink atom with arrow).} \label{fig:nvo}
\end{center}
\end{figure*}

The role of oxygen vacancies is simulated through the removal of one oxygen atom within our supercell (120 atoms) and allowing the internal geometry to relax. Since the oxygen within \ahs has an oxidation state of O$^{2-}$, oxygen vacancy is an n-type dopant, which donates two electrons to the system. The band structure of hematite with one neutral oxygen vacancy (Figure \ref{fig:vo}) displays the creation of three defect states within the band gap $\alpha$, $\beta$, $\gamma$ in agreement with previous theoretical work \cite{theoryvo}. The corresponding projected density of states is included in SI figure 2. Defect states are considered to be lozalized states (over a few nearest neighbor atoms -- see SI figure 7) and have no dispersion as seen by the band structure and density of states. We found the two electrons are spontaneously ionized from oxygen vacancy site and form localized small polarons at nearby Fe ion (no charge density appears at the center of the vacancy, unlike the case of SrTiO$_3$ \cite{srtio3}). The $\beta$ state in the band structure corresponds to the two polaron states (one spin up and one spin down) in which the extra electrons from \vo occupy (the wave functions are shown in Figure \ref{fig:vo}). They are $d_{x^2-y^2}$ type orbitals of the two nearest neighbors of the four-coordinated \vons. The $\alpha$ state is a perturbed valence state of O $2p$ orbitals surrounding the site at which the extra electrons go, while the $\gamma$ state is a perturbed conduction state which is unoccupied and corresponds to $e_g$ orbitals of the next nearest Fe neighbors of \vons. Overall, the oxygen vacancy did not modify the band edge positions and band dispersion of the pristine systems, except introducing isolated defect levels in the band gap. Therefore, we do not expect it will affect the main optical absorption spectra (as the defect states have low density of states) but it may affect the transport properties of the system, i.e. the carrier concentration or mobility as we will discuss later.

\subsubsection*{N-doped + Oxygen Vacancy}

Atomic nitrogen has been found to be a promising dopant to improve visible light absorption properties of metal oxides, such as in TiO$_2$ \cite{n:tio2} and BiVO$_4$ \cite{ping}. Recently, we found N doping in the presence of oxygen vacancies can also improve the carrier mobility of BiVO$_4$, where the small polaron hopping barrier was lowered. Yet, N doping in \ahs with/without oxygen vacancies has not been studied both theoretically and experimentally to our best knowledge. We will firstly discuss the effects of N doping and then discuss the combination effect of N doping with oxygen vacancies. \\

For N-doped hematite we replaced one O in our system with N, a p-type dopant since N has one fewer valence electron than O. This substitution results in a deep defect state created corresponding to an unoccupied N $2p$ orbital and results in slightly perturbed valence band states (See SI Figure 3 for band structure and projected density of states.) The deep defect state introduced by N doping has minimal overlap with the O $2p$ states, which may introduce defect state involved transitions in the optical spectra but will not affect the main absorption edge. In the meanwhile, the VBM in N-doped hematite is raised by 0.1 eV due to the hybridization between N and O 2p states, and interestingly, the indirect band gap for the pristine \ahs becomes a direct band gap after N doping. This could both shift the absorption edge to lower energy slightly and also enhance the absorption coefficient at the same energy range compared to the pristine case.


\begin{figure*}
\begin{center}
\includegraphics[keepaspectratio=true,scale=1]{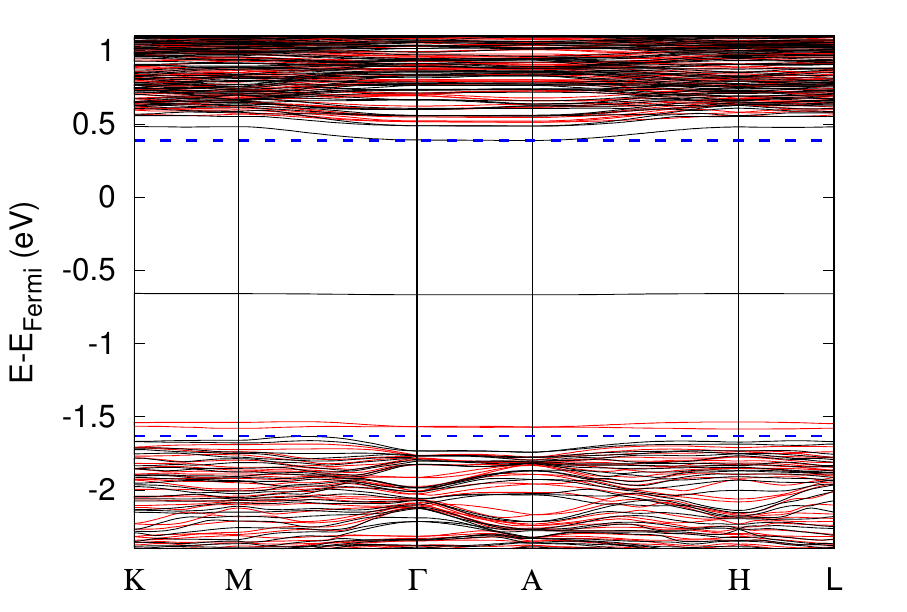} 
\includegraphics[keepaspectratio=true,scale=0.1075]{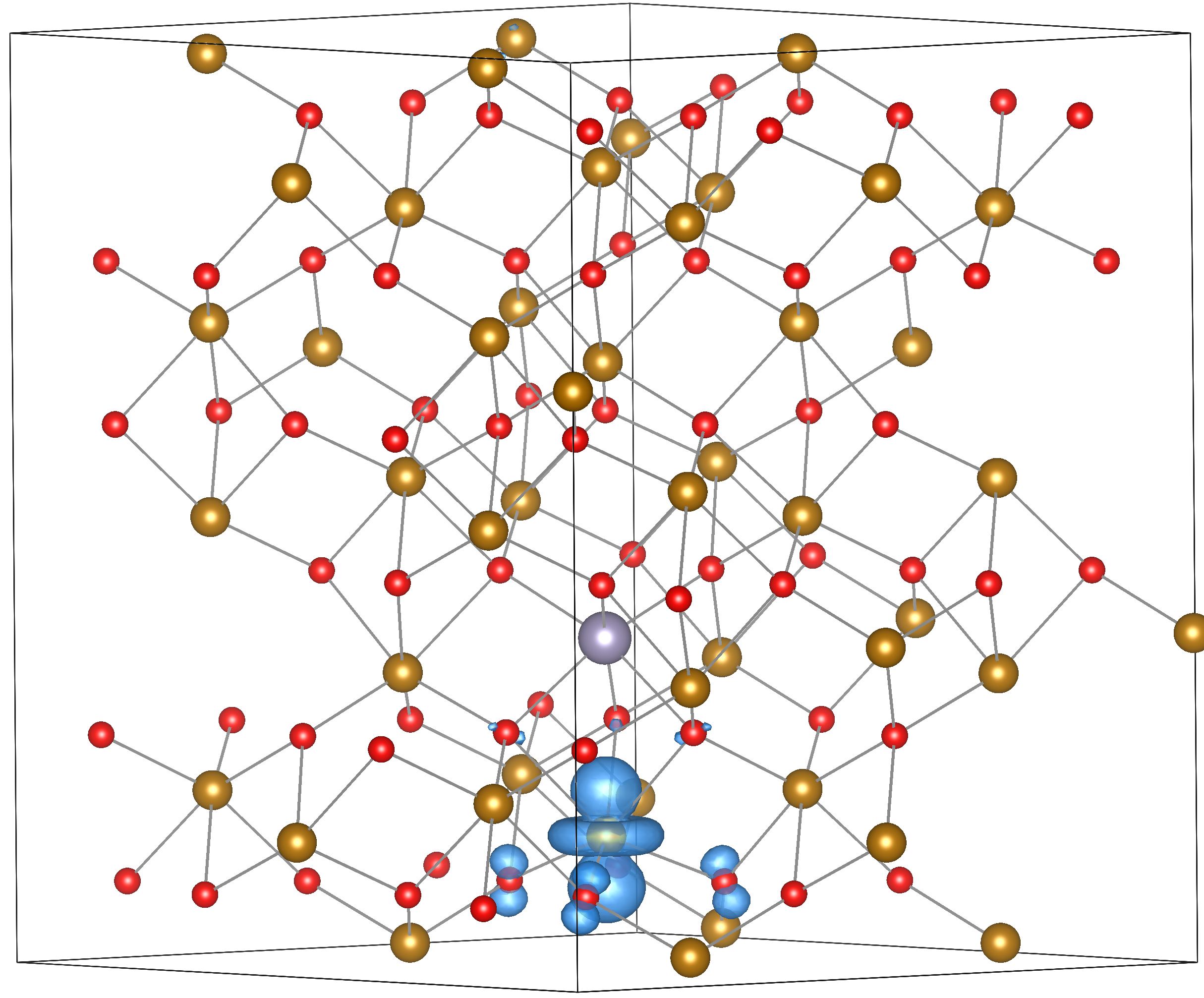}
\caption{(Top) The band structure of Sn:Fe$_2$O$_3$ with a perturbed conduction band edge of Fe $3d$ states creating a reduced gap of 2.0 eV (see SI figure 6 for details). There is a single isolated defect introduced in the gap which corresponds to a small polaron formed at Fe site near Sn. (Bottom) Isosurface plot (blue) of the highest occupied orbital present in Sn:Fe$_2$O$_3$, a small polaron formed directly below Sn (grey atom).} \label{fig:sn}
\end{center}
\end{figure*}


Next, we considered nitrogen doping coupled with oxygen vacancy in both a (1N:1\vons) ratio and a (2N:1\vons) charge-balanced ratio. In the (1:1) doping regime the overall effect is still n-type yet one of the polaron states is absorbed by the nitrogen into the previously unoccupied N $2p$ state (Figure \ref{fig:nvo}). We found that the defects prefer to be close-by, rather than far away, with $E_{near}-E_{far}\approx -0.67$ eV, with only small variations in the $E_{near}$ based on N and \vo orientation and distance. Tested configurations had the distance between \vo and N in the `near' case as about $\sim\,$2.5-3.0 \AA $\ $and $\sim\,$7-10 \AA $\ $in the `far' case. The resulting electronic structure of N+\vons :Fe$_2$O$_3$ is shown in Figure \ref{fig:nvo} and displays a slightly reduced band gap of 2.1 eV (see SI Figure 4 for details). Again, as in N-doped hematite, we see a shift at the VBM of N+\vons :Fe$_2$O$_3$, resulting in a direct band gap introduced from N-doping. \\


\begin{table}[t!]
\begin{center}
\begin{tabular}{p{1.9cm}p{1.9cm}p{1.9cm}p{1.9cm}}
 \hline
 \multicolumn{4}{c}{ 2N:1\vo Hematite configurations \vspace{0.5mm}} \\ 
 \hline
 $\beta$ polarons & N-N ({\AA}) & N-Fe-N ($^{\circ}$)& $\Delta E_{tot}$ (eV) \\
 \hline
 2  & 2.4025 & 74.578 & 0.44 \\
 1  & 2.7482 & 86.841 & 0.23 \\
 0  & 3.8606 & 151.269 & \hspace{1mm} 0 \\
 \hline
\end{tabular}
\end{center}
\caption{Summary of 2N:1\vo configurations, displaying the relation between number of $\beta$ polarons from \vo formed, distance between N (N-N ({\AA})), atomic angle (N-Fe-N ($^{\circ}$)) and the total energy difference ($\Delta E_{tot}$) with the most stable system (the configuration with 0 $\beta$ polarons formed).} \label{table:n2vo}
\end{table}


 We also tested nitrogen doping coupled with oxygen vacancy in a (2:1) ratio and we found three possible electronic configurations based on the orientation of the defects within the lattice (in each case they are attached to the same Fe ion). These configurations are nicely characterized by the number of $\beta$ type states ($\beta$ polarons) created by \vo. A summary of these configurations is collected in Table \ref{table:n2vo}. We found that if all three defects are as close as possible (with N-N distance 2.40\AA) then both small polarons are formed (the number of $\beta$ polarons is 2 as shown in Table \ref{table:n2vo}). If all three are still close yet the nitrogen atoms are on opposite sides of \vo (with N-N distance 3.86\AA)then both small polarons are absorbed by the nitrogen (the number of $\beta$ polarons is 0 as shown in Table \ref{table:n2vo}; see detailed structures in SI figure 8). Lastly, it is possible to form a configuration somewhere in between these two states and have only one small polaron state formed (the number of $\beta$ polarons is 1 as shown in Table \ref{table:n2vo}). Ultimately, the most stable configuration for the (2:1) ratio is to have the nitrogen on opposite sides of \vo so that they may absorb the two electrons of \vo with minimal interaction between nitrogen sites. This configuration is minimal in energy because the absorption of these extra electrons allows the nitrogen to achieve filled valence bands while keeping their repulsive electrostatic interaction at a minimum. In any of these configurations, several defect states are still present within the band gap (see the corresponding band structure and projected density of states in SI figure 5).  Although both electrons donated from \vo are absorbed by the nitrogen atoms in the most stable configuration of 2N:1\vons , these dopants still introduced extra defect states inside the gap. This is unlike the case of BiVO$_4$, where the charge balanced N doping with oxygen vacancy results in the shift of valance band edges but no isolated defect states formed in the band gap \cite{ping}. Meanwhile, the valence band edge shifts up by 0.2 eV in this case resulting in a reduced band gap size of 2.0 eV, and is also accompanied by an indirect to direct 
 gap transition.

\subsubsection*{Sn-doped}

Sn substitution has been shown experimentally to be a promising dopant for improving photocurrents in \ahs photoanodes \cite{yatsn}. However, similar to the case of oxygen vacancies, it is unclear whether this is because the bulk carrier conductivity has been improved. Sn is a n-type dopant since the substitution Fe$^{3+}\rightarrow$ Sn$^{4+}$ donates an electron into the system. The atomic number of tin is nearly double that of iron (50:26) yet the atomic radius of Sn$^{4+}$ (83.0 pm) is only slightly larger than that of Fe$^{3+}$ (78.5 pm). Since the Fe ions in hematite have a high spin configuration of half-filled $3d$ orbitals, the main modification of the electronic structure upon the introduction of Sn substitution is that Sn has no unpaired spin which creates a hole in the magnetic ordering of the anti-ferromagnetic arrangement of \ah . Additionally, Sn valence orbitals are less localized than Fe $3d$ orbitals, which results in the excess electron easily moving away from Sn and localizing on a nearby Fe ion. Therefore, the excess electron contributed from Sn substitution forms a small polaron at the Fe site located either directly below or directly above Sn (depending on the location of Sn in the lattice, Figure \ref{fig:sn}). Alignment of Fe 3s semi-core states between the pristine and doped systems reveals the conduction band edge down-shift of unoccupied Fe $3d$ orbitals in Sn-doped hematite resulting in a slightly smaller band gap (2.0 eV) than pristine \ahs (see Figure \ref{fig:sn} for the band structure and SI Figure 6 for the projected density of states). Overall, Sn doping causes one small polaron defect state in the gap and perturbs the states close to the band edges, unlike the case of oxygen vacancies which only results in several isolated defect states in the gap with no perturbation of band edges.

\section*{Formation Energy of Charged Defects}

To understand whether the defects can be ionized easily at the room temperature (and therefore contribute to the carrier concentrations), we calculated thermodynamical charge transition levels $\epsilon_{0/+1}$ and $\epsilon_{+1/+2}$ of the doped systems considered. The charge transition levels are defined to be the Fermi level where the formation energies of two different stable charged states are equal. The formation energy is given by
\begin{equation}
E_{for}(defect,q) = \Delta E_{tot} +\lambda\,\mu_x+q\,\varepsilon_{F}+\Delta_q \label{eq:ef}
\end{equation}
where $\Delta E_{tot} = E_{tot}(defect,q)-E_{tot}(pristine)$ is the difference in total energy for the charged defect system to that of the pristine system, $\lambda\,\mu_x$ is the chemical potential of any removed or added species, $q\,\varepsilon_{F}$ is the electron reference with $\varepsilon_{F}$ measured from the valence band maximum (of the pristine system) and $\Delta_q$ is the charged cell correction to correct the spurious interactions between charges in periodic images. The charged cell correction is computed as discussed in the method section \cite{ping2}, while the chemical potentials of elements Fe, Sn, O and N were estimated as the total energy per atom of the natural occurrence of the element. In this paper we focus on the charge transition levels of the defects which will not depend on the chemical potential of elements (e.g. oxygen poor or rich conditions).\\


\begin{figure}[t!]
\begin{center}
\includegraphics[keepaspectratio=true,scale=0.72]{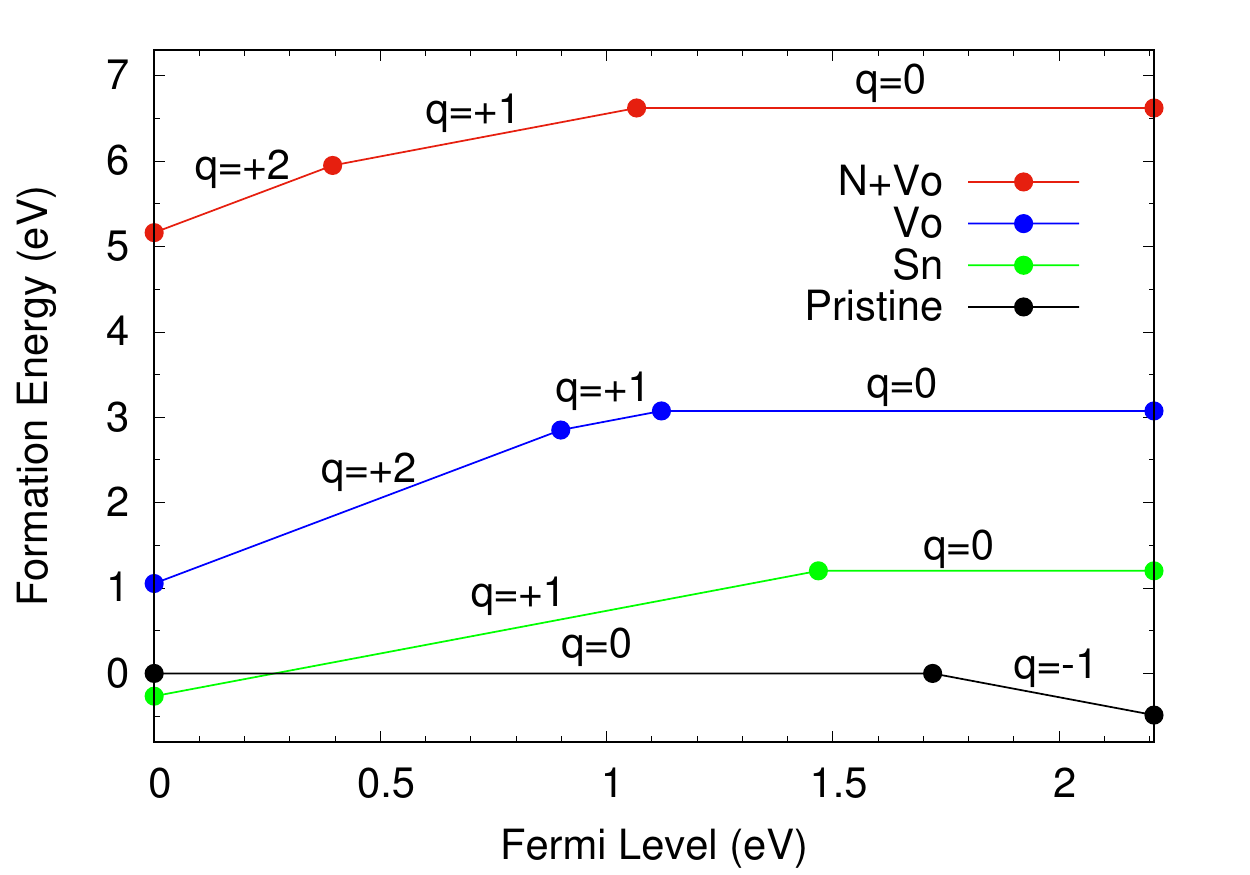} 
\end{center}
\caption{Formation energy displaying the most stable charge states of the different doped systems considered with respect to the Fermi level. The zero Fermi level is the VBM of the pristine system.} \label{fig:ef}
\end{figure}



\begin{table}[t!]
\begin{center}
\begin{tabular}{p{2.5cm}p{2.5cm}p{2.5cm}}
 \hline
 \multicolumn{3}{c}{ Ionization Energies in Doped Hematite \vspace{0.5mm}} \\ 
 \hline
 Doping & (+1/+2) eV & (0/+1) eV\\
 \hline
 \vons  & 1.31  & 1.09 \\
 N+\vons & 1.82 & 1.14 \\
 Sn & -- & 0.74 \\
 \hline
\end{tabular}
\end{center}
\caption{Corresponding ionization energies obtained from the energy difference between the charge transition level to the conduction band for different doping cases.} \label{table:ef}
\end{table}

\begin{figure}[t!]
\begin{center}
\includegraphics[keepaspectratio=true,scale=0.72]{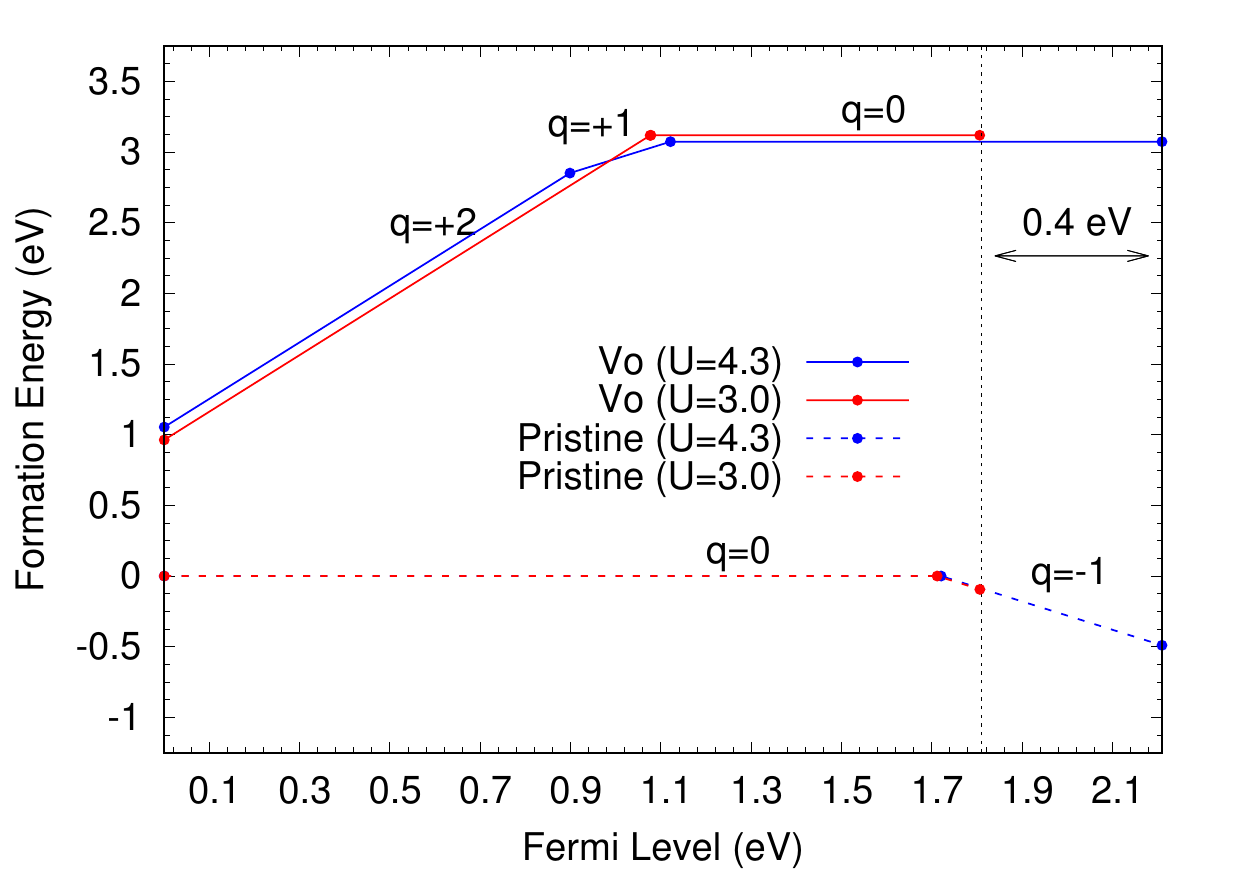} 
\end{center}
\caption{Formation energy plot for pristine hematite and hematite with \vo with different U parameters, 3.0 and 4.3 eV. The reference zero of the Fermi level is the VBM of the prisitine system. 
The charge transitions still occur at the similar Fermi level with respect to the VBM in the respective systems, however, a smaller U parameter results in a smaller band gap. In particular $E_{gap}=1.81$ eV in the $U=3.0$ eV system, 0.4 eV smaller than $E_{gap}=2.21$ eV in the $U=4.3$ eV system.
As a result, the defect ionization energy is lowered by 0.4 eV at U= 3.0 eV compared with U=4.3 eV.} \label{fig:efu}
\end{figure}

From Eq. \ref{eq:ef} we obtained the charged defect formation energies and charge transition levels for the doped systems as displayed in Figure \ref{fig:ef}. The ionization energies of these states are given by the difference of the conduction band minimum 2.21 eV (of the pristine system) to the Fermi level at which the charge transitions occur, summarized in Table \ref{table:ef}. We see that \vo introduces deep defects into the system with large ionization energies 1.31 eV (+1/+2) and 1.09 eV (0/+1), consistent with previous theoretical work which reported 1.08 eV for (0/+1) \cite{theoryvo}. Meanwhile we see that N+\vo deepens the (+1/+2) transition by 0.51 eV yet does not significantly change the (0/+1) ionization energy. Sn has the lowest ionization energy of 0.74 eV (0/+1), which is larger than previously reported values by 0.22 eV \cite{groupIV}; yet, they also reported a smaller band gap (2.15 eV) than ours (2.21 eV) due to a lower U value (4 eV). It is important to note that these ionization energies indirectly depend on the Hubbard U parameter. Specifically, a lower U parameter will lower the conduction band minimum position and hence result in lower defect ionization energies. We find that the charge transition levels are not expected to change much with respect to the valence band maximum, yet the band gap can have a range from 1.8 eV to 2.2 eV for a Hubbard U of 3.0 eV to 4.3 eV (see Figure \ref{fig:efu}). Nonetheless, the overall defect property does not change qualitatively, i.e. the oxygen vacancy has an ionization energy ranging from 1.09 eV (at U=4.3 eV) to 0.79 eV (at U=3.0 eV), which is still a deep impurity compared with the ionization energy of a small polaron in the prisitine hematite (0.49 eV at U=4.3 eV and 0.09 eV at U=3.0 eV.)


We note that although the absolute Sn-doping ionization energy is still large (0.74 eV), if we compare its ionization energy with the ionization energy of one small polaron in a pristine system (0.49 eV; black line in Figure \ref{fig:ef}), the difference is only 0.25 eV. This implies that it takes 0.25 eV to excite the small polaron from a defect-bound polaron to an unbound polaron (or `free' polaron), which can be considered as a shallow impurity. And because electrons naturally form small polarons even in the pristine \ahs and it takes 0.49 eV to ionize the electrons from self-trapped to free electrons in the conduction bands, absolute `shallow' n-type defects (where the ionization energy relative to the CBM is at the order of $kT$) can hardly form in \ah . In particular,this is evident for the ionization energies of \vo and N+\vo which are still $\sim\,$0.6-0.7 eV away from the ionization energy of one small polaron in the pristine system. Therefore, even when considering the ionization energy of a small polaron in the pristine system, the polarons formed in these systems (\vo and N+\vons ) are still difficult to ionize and contribute to carrier concentrations at room temperature, unlike the case of Sn doping. \\

\section*{Small Polaron Transport in Pristine and Defective Hematite}

\begin{figure}[t!]
\begin{center}
\includegraphics[keepaspectratio=true,scale=0.69]{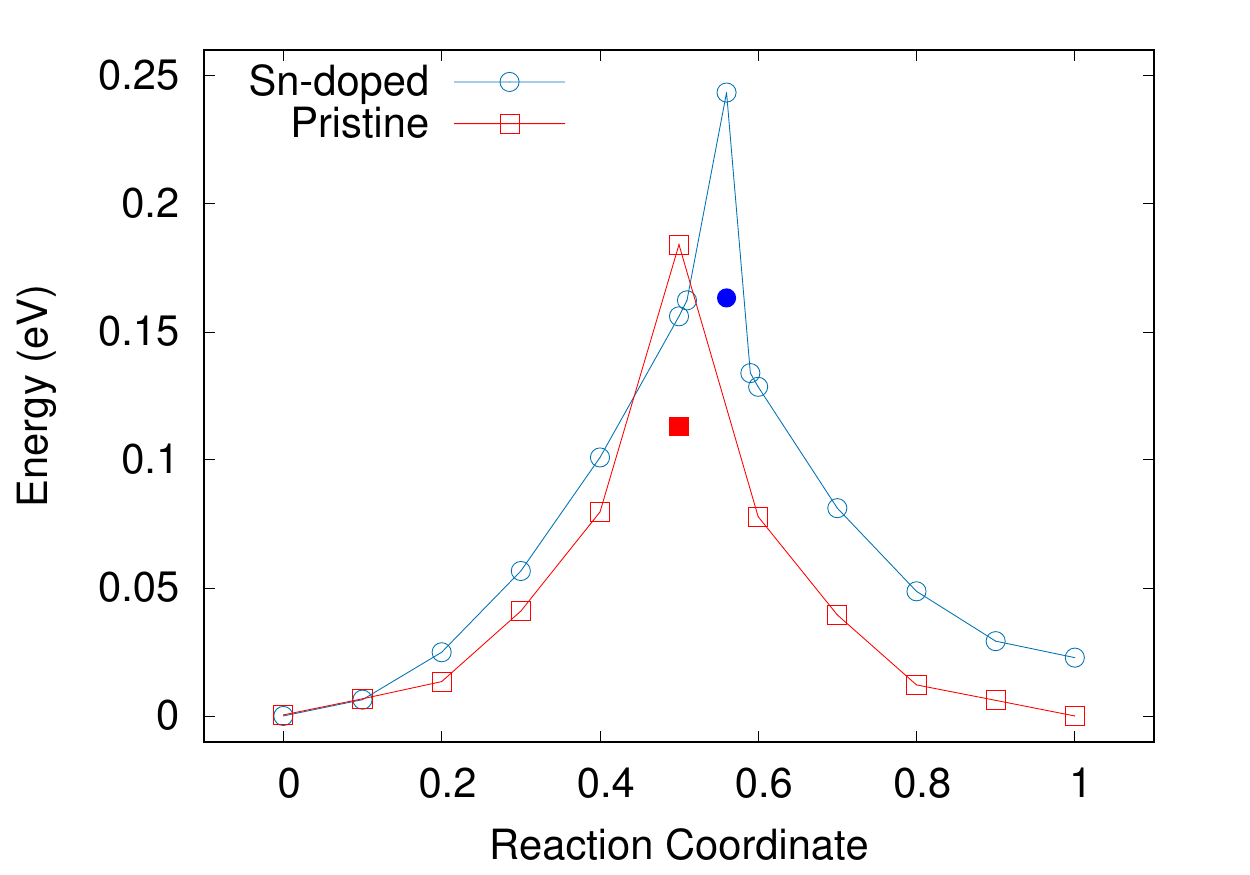} 

\vspace{2mm}

\includegraphics[keepaspectratio=true,scale=0.175]{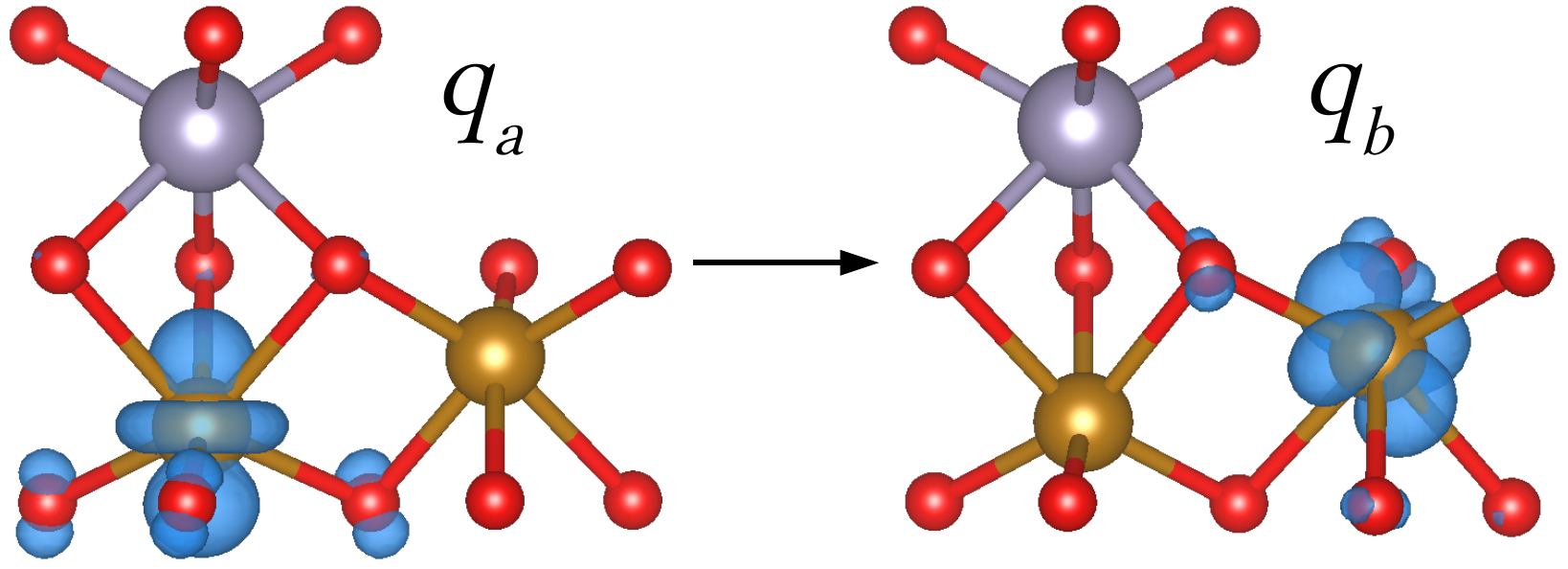}
\end{center}
\caption{(Top) Small polaron hopping barrier in the pristine and Sn-doped systems where the solid square and dot indicate the activation energy. (Bottom) Schematic of polaron hopping in the Sn-doped system (left is the initial $a$ configuration and $b$ is the final configuration).} \label{fig:transport}
\end{figure}

From the ionization energy calculations above, we have found Sn is a promising dopant for carrier concentration improvement. As the carrier conductivity depends on both the carrier concentration and mobility, in this section we will discuss whether Sn doping improves the carrier mobility as well. Previous work \cite{adiabatic} found that in pristine \ahs the coupling of small polaron hopping sites (V$_{AB}$) is large relative to the reorganization energy $\Delta$G* (V$_{AB} > \Delta$G*/4), implying the small polaron hopping conduction of \ahs is in the adiabatic regime. We then computed the small polaron hopping activation energy of pristine and Sn-doped hematite using a linear extrapolation technique, which includes the coupling V$_{AB}$ implicitly. We start from the initial configuration $q_{a}$ where the small polaron is located on a Fe ion and then using geometrical techniques described in the method section determine a configuration $q_{b}$ in which the small polaron is located on a nearest neighbor Fe ion in the same $ab$-plane. Then, in accordance with Eq. \ref{eq:ex} we linearly extrapolated from $q_{a}$ to $q_{b}$, allowing the electron density of the system to equilibrate at each step but keeping the geometry $q_{x}$ fixed. 
\begin{equation}
q_{x} = q_{a}(1-x)+x\,q_b \label{eq:ex}
\end{equation}
The computed total energies (referenced to the starting configuration) as a function of reaction coordination are shown in Figure \ref{fig:transport}. The peak of the barrier is the saddle point between the two hopping sites and the true activation energy $E_{A}$ is obtained from relaxing the geometry from this point. In the pristine system the saddle sits at the $q_{0.5}$ reaction coordinate due to the symmetry of two hopping centers and in Sn:Fe$_2$O$_3$ it sits at the $q_{0.56}$ reaction coordinate. The latter relates to the different distances from the small polaron center to  the Sn defect center.\\


For the pristine system we obtain an activation energy of $E_{A,Fe} = 0.11$ eV, consistent with previously calculated results of 0.11 eV \cite{adiabatic} and 0.13 eV \cite{transport}. In the Sn-doped system we computed an activation energy of $E_{A,Sn}=0.16$ eV. This higher activation energy is quite significant when compared with $kT$ at room temperature ($k$ is the Boltzmann constant and $T$ is temperature). In particular the electron transfer rate is proportional to the exponential of the activation energy over $kT$ as shown in Eq.\ref{eq:mob}. 
\begin{equation}
\tau = A e^{-E_{A}/kT} \label{eq:mob}
\end{equation}
where $\tau$ is the electron transfer rate and $A$ is the prefactor which depends on the number of nearest neighbors and the attempt frequency. 
Considering that both systems have 3 nearest neighbors and Sn disrupts the geometry only slightly,  the prefactor $A$ of these two systems should be on the same order of magnitude. The hopping mobility is related to the electron transfer rates through the Einstein relation and one dimensional random walk i.e. $\mu=eR^2\tau/(2kT)$, where R is the electron transfer distance and $\tau$ is the electron transfer rate\cite{transport,einstein}. Therefore we can determine the ratio of the mobility in these two systems to be solely dependent on the difference of their activation energies. 

\begin{equation}
\frac{\mu_{Sn}}{\mu_{Fe}} \simeq e^{-(E_{A,Sn}-E_{A,Fe})/kT} \label{eq:mobratio}
\end{equation}
Eq.\ref{eq:mobratio} gives a ratio $\mu_{Sn}/\mu_{Fe} = 0.14$, which implies a decrease in carrier mobility in Sn:Fe$_2$O$_3$. We note that the effect of defects on the carrier mobility depends on the distance to the defect center, i.e. the hopping barriers between the pristine and Sn-doped systems will be similar when the hopping centers are far away from the defects (Sn). Therefore what we estimate here is the lower bound of carrier mobility at presence of Sn.\\

We also investigated the hopping barriers at presence of \vo and N+\vo by using similar procedures; yet we found the small polaron centers are unstable away from the defects (the electrons tend to localize closest to the defect centers, despite any applied local distortion at the positions far away from the defects.) This again indicates these defects in bulk \ahs are deep and extra electrons from  \vo  and N+\vo  tend to form small polarons tightly bounded to the positive charged defect centers.


\section*{Conclusions}

In summary, this work discussed the effects of defects on the small polaron formation, ionization and hopping transport properties in bulk \ahs through first-principles calculations, where the choice of dopants is inspired by recent experimental and theoretical work\cite{yatvo,yatsn,yatti,ping}. Our calculations of the electronic structure of pristine and doped \ahs show that the small polarons will naturally form at Fe ions if there are any excess electrons (unless there are dopants with stronger correlated electrons available than the $3d$ electrons of Fe). These small polarons can be tightly bounded to the defect centers (strongly preferred to locate at Fe closest to the defects) or possibly conduct through thermally activated small polaron hopping, depending on the corresponding ionization energies. 
 
Electronic structure calculations show the defects we investigated here have different effects on the band structures: for \vons , only deep defect levels are introduced in the band gap and the bulk band structure did not change compared with the pristine hematite. For the cases of N,  (1N:1\vons ) and (2N:1\vons ) doping, we observe 1) defect levels appear in the band structure; 2) the valence band maximum shifts up by 0.1-0.2 eV, accompanied by an indirect to direct gap transition, which could improve the visible light absorption at the same energy range as before, but also lower the absorption spectra edge than the pristine hematite. The enhancement of visible light absorption may be further confirmed by computing absorption spectra for these doped systems in the future work. For Sn doping, we found related defect states in the gap and also a band gap reduction by 0.2 eV by lowering of the conduction band position. 

Formation energy as a function of Fermi level reveals most stable charge configurations of the several doped systems of hematite considered: \vons, N+\vo and Sn. From our defect formation energy calculations, we see that \vo and N+\vo bind the small polarons strongly with relatively large ionization energies. In the Sn-doped system we see a much lower ionization energy (it takes 0.25 eV from a defect-bound polaron to a "free" polaron) compared to the pristine system. This shows Sn doping could contribute to the carrier concentrations at room temperature. We note that due to the spontaneous formation of small polarons even in pristine systems (with an ionization energy of 0.49 eV to become free electrons in CBM), an absolute shallow n-type impurity can hardly form in \ahs since this would require an ionization energy relative to CBM more comparable to $kT$ at room temperature ($\sim\,$0.026 eV). 

Calculations of the small polaron hopping activation energy were conducted in pristine and Sn-doped hematite and we found that the introduction of Sn may lower the hopping mobility due to a higher hopping activation barrier. However, the improved carrier concentration and possibly improved light absorption  by Sn doping can improve the overall photocurrents as observed experimentally. We note that although experimentally they also found \vo  improved the photocurrents of \ahs \cite{yatvo,yatsn} (which is not supported by our theoretical results that \vo is a deep impurity and does not modify the band structure of the pristine system except introducing defect bands), the measurements are based on an average of bulk and surface defects, at the presence of electrode/electrolyte interfaces. In order to further elucidate the roles of \vons, N+\vo and Sn doping in \ahs for solar water splitting applications, the ionization energies of charged defects at surfaces as well as the interaction between small polarons, defects and water at the \ah /water interface needs to be investigated in the future work.


\section*{Acknowledgments}
The authors thank Yat Li and Feng Wu for useful discussions. The authors acknowledge the Texas Advanced Computing Center (TACC) at The University of Texas at Austin for providing high performance computing resources that have contributed to the research results reported within this paper \cite{stampede}. The authors also acknowledge resources from the Center for Functional Nanomaterials, which is a U.S. DOE Office of Science User Facility, at Brookhaven National Laboratory under Contract No. DE-SC0012704. T. J. Smart acknowledges full support from the Department of Education GAANN fellowship.

\bibliography{references.bib}
\bibliographystyle{apsrev}

\onecolumngrid


\section*{Supplementary material}

## Supporting Information for "Effect of Defects on the Small Polaron Formation and Transport Properties of Hematite from First-Principles Calculations"

### $Fe_2O_3$ Band Structure

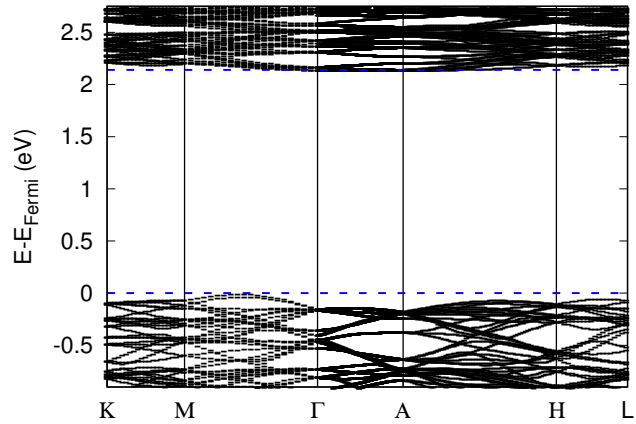

**Supplemental Figure 1.** The band structure of $\alpha$-$Fe_2O_3$ (with a 2×2×1 supercell) with symmetric points $K = (1/3, 1/3, 0)$, $M = (1/2, 0, 0)$, $\Gamma = (0, 0, 0)$, $A = (0, 0, 1/2)$, $H = (1/3, 1/3, 1/2)$ and $L = (1/2, 0, 1/2)$. The blue dashed lines show the positions of conduction band minimum (CBM) which is located between $\Gamma$ and $A$, and valence band maximum (VBM) which is located between M and $\Gamma$.

### $V_O$:$Fe_2O_3$ PDOS and Band Structure

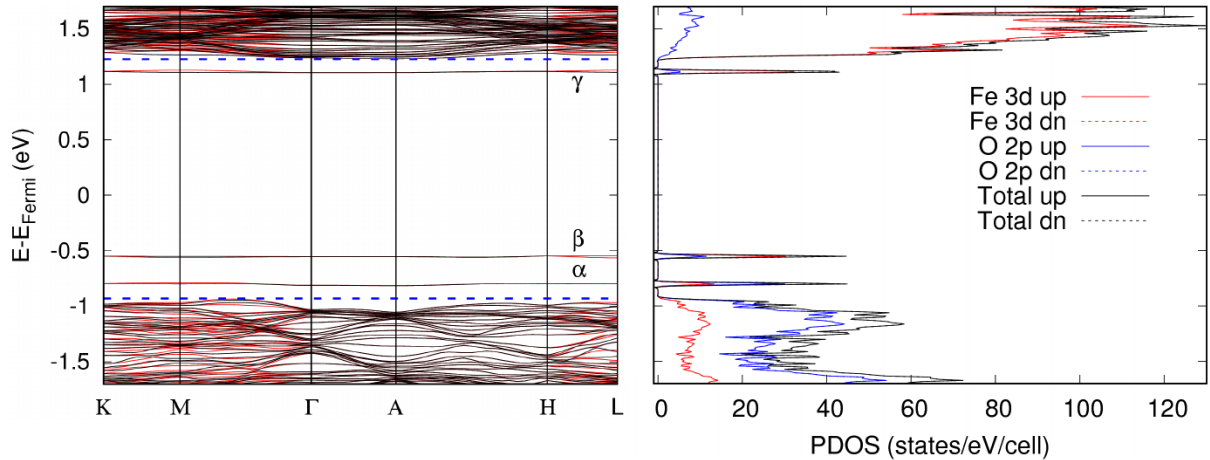

**Supplemental Figure 2.** The band structure of $V_O$:$Fe_2O_3$ accompanied by the projected density of states. The PDOS shows 1) clearly isolated defects $\alpha$, $\beta$ and $\gamma$ which form in the gap, 2) the system remains spin symmetric and 3) neither of the band edges are not notably perturbed from $V_O$.

**N:Fe$_2$O$_3$ PDOS and Band Structure**

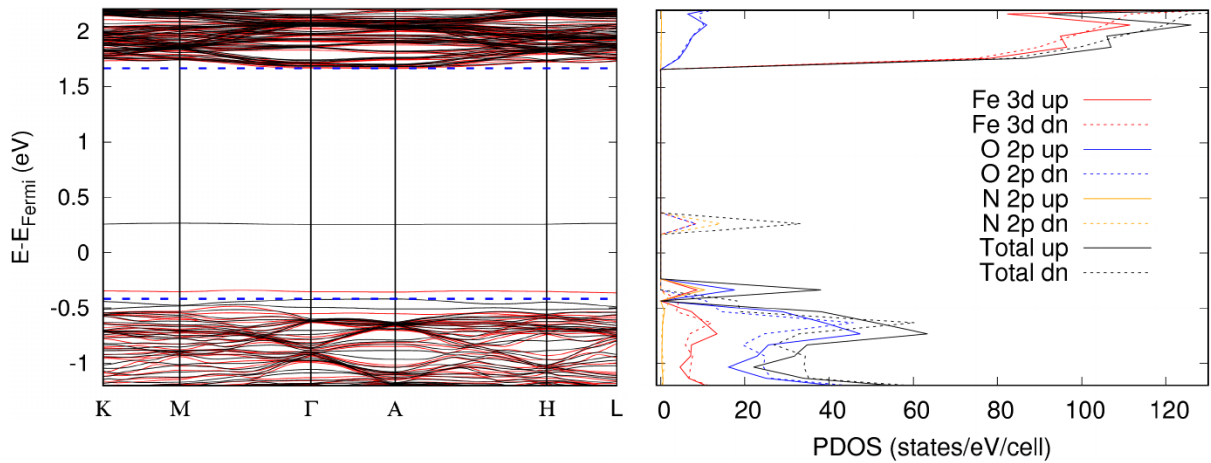

**Supplemental Figure 3.** The band structure of N:Fe$_2$O$_3$ accompanied by the projected density of states. There are two defect bands which correspond to an unoccupied spin down N 2$p$ state and an occupied spin up N 2$p$ state. Notably the valence band edge is perturbed by the introduction of N with minimal overlap of N states (as seen from PDOS). This results in not only a slight reduction of the band gap (2.1 eV) but also a indirect to direct gap transition.

**N+V$_O$:Fe$_2$O$_3$ PDOS and Band Structure**

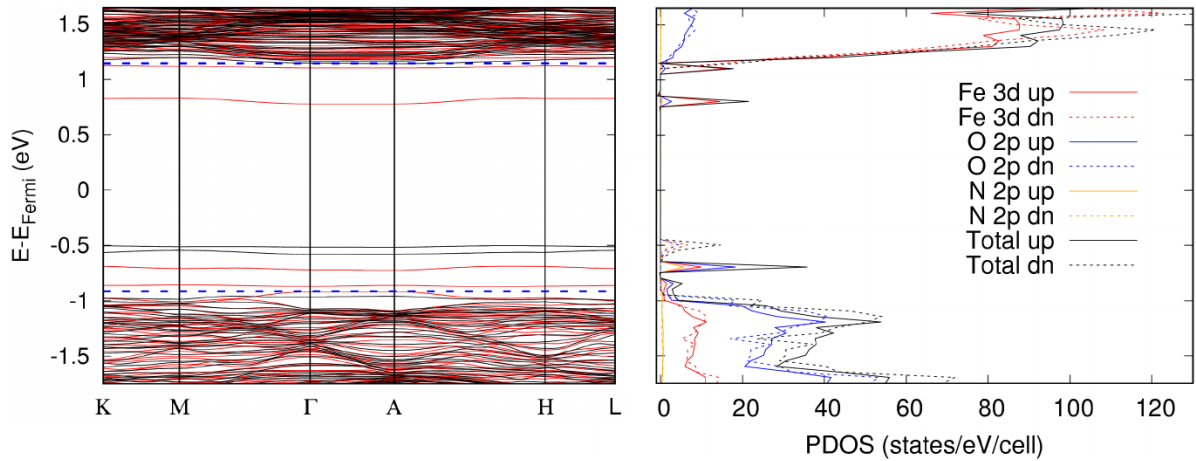

**Supplemental Figure 4.** The band structure of N+V$_O$:Fe$_2$O$_3$ accompanied by the projected density of states. Despite the presence of V$_O$, this case is similar to the case of N:Fe$_2$O$_3$ in that the VBM is shifted resulting in a reduced gap (2.1 eV) and a newly formed direct gap transition.

## 2N+V$_O$:Fe$_2$O$_3$ PDOS and Band Structure

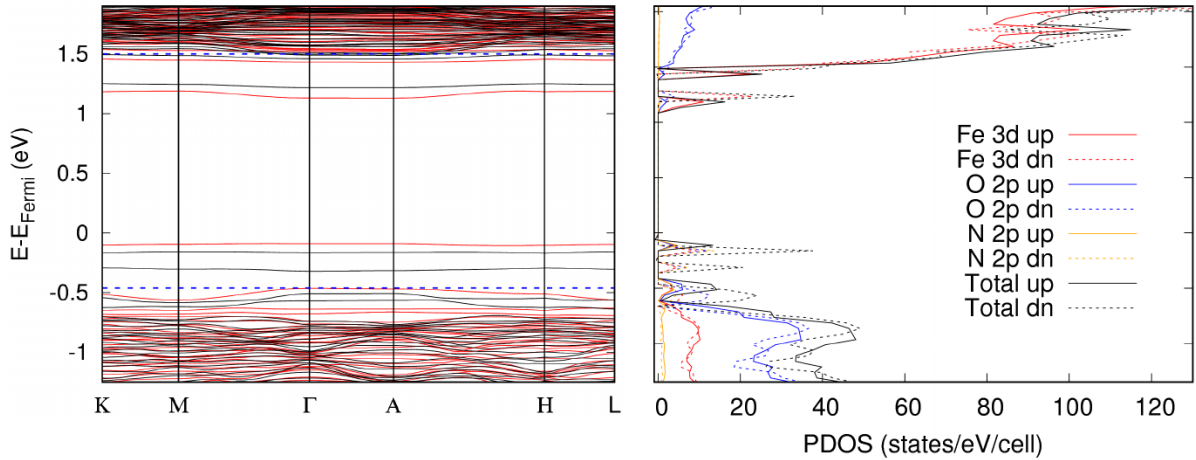

**Supplemental Figure 5.** The band structure of 2N+V$_O$:Fe$_2$O$_3$ accompanied by the projected density of states. The addition of a second nitrogen results in an even greater perturbation of the VBM and a reduced gap of 2.0 eV. The unoccupied defects lying at about $E - E_{Fermi} \sim 1.2$ eV correspond to the two unoccupied $\beta$ polarons which are absorbed into occupied N 2$p$ states which compose the occupied defects directly below the Fermi level.

## Sn:Fe$_2$O$_3$ PDOS and Band Structure

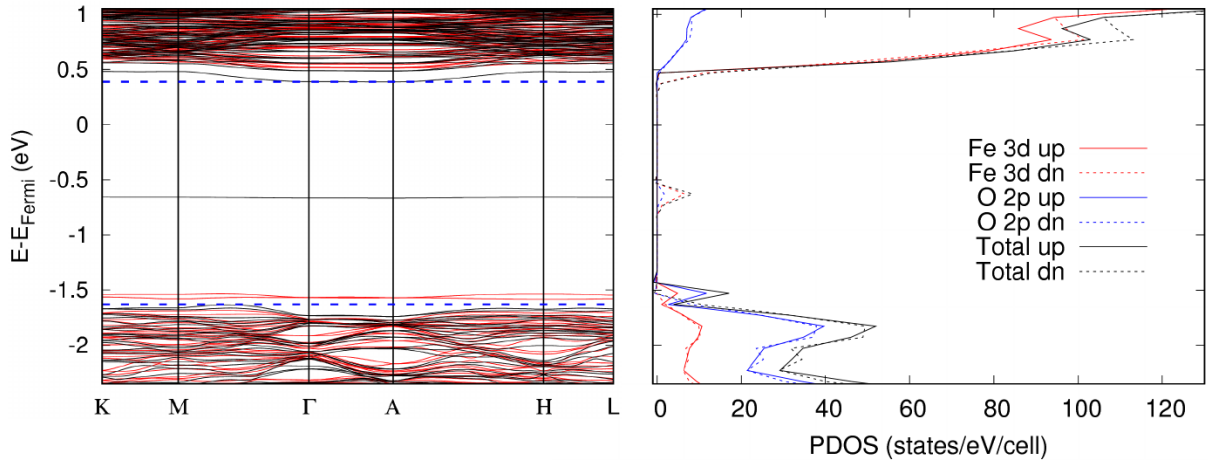

**Supplemental Figure 6.** The band structure of Sn:Fe$_2$O$_3$ accompanied by the projected density of states. The PDOS shows 1) clearly isolated defect in the center of the gap which corresponds to a small polaron, 2) two defect states from the VBM are formed which correspond to O 2$p$ states surrounding the small polaron site and 3) spin down conduction band states are perturbed from Sn replacing a down Fe ion site resulting in a band gap reduction by 0.2 eV.

**Defect vs. Band States in Sn:Fe₂O₃**

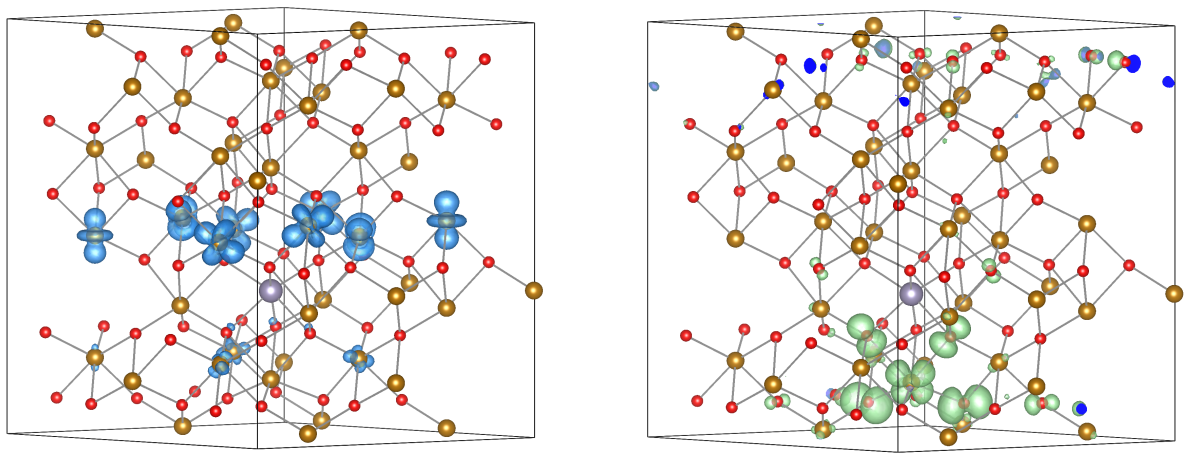

**Supplemental Figure 7.** Throughout this work we have distinguished between defect bands and valence/conduction band states in order to understand the influence of defects on band gap properties of hematite. The above figure is an example in the case of Sn:Fe₂O₃. (left) A perturbed conduction band state considered to be the CBM of Sn:Fe₂O₃ and (right) a defect state directly above the valence band not considered to be in the VBM of Sn:Fe₂O₃. Similar reasoning is used in other cases, with the rule of thumb being whether or not the state is localized on only a few sites.

**2N+V_O:Fe₂O₃ Configurations**

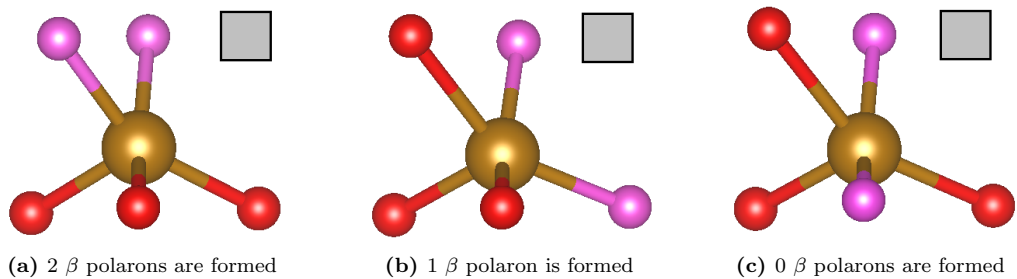

**(a)** 2 $\beta$ polarons are formed  **(b)** 1 $\beta$ polaron is formed  **(c)** 0 $\beta$ polarons are formed

**Supplemental Figure 8.** The local structure of the three most stable configurations of 2N+V_O doped hematite. Significantly different electronic energies and electronic structures result from these slightly different orientations of V_O and N (gold:Fe, red:O, pink:N, grey:V_O). Structure (c) is the most stable where 1) the N have absorbed both $\beta$ polarons, 2) the negatively charged N are close in proximity to V_O, 3) the N are on opposite sides of Fe (as far as possible while remaining close to V_O).

| 2N:1V_O Hematite configurations | | | |
|---|---|---|---|
| $\beta$ polarons | N-N (Å) | N-Fe-N (°) | $\Delta E_{tot}$ (eV) |
| 2 | 2.4025 | 74.578 | 0.44 |
| 1 | 2.7482 | 86.841 | 0.23 |
| 0 | 3.8606 | 151.269 | 0 |

Supplemental Table I: Summary of 2N:1V_O configurations, displaying the relation between number of $\beta$ polarons from V_O formed, distance between N (N-N (Å)), atomic angle (N-Fe-N (°)) and the total energy difference ($\Delta E_{tot}$) with the most stable system (the configuration with 0 $\beta$ polarons formed) as shown in table I.


\end{document}